\documentclass[traditabstract]{aa}    

\usepackage{graphicx}
\usepackage{txfonts}
\usepackage[colorlinks=true,citecolor=blue]{hyperref}
\usepackage{graphicx}
\usepackage{times}
\usepackage{natbib}
\usepackage{color}
\usepackage{multirow}
\usepackage{todonotes}
\usepackage{appendix}
\usepackage{array}
\usepackage{verbatim}

\begin{document} 

\title{On the redshift evolution of the baryon and gas fraction in simulated groups and clusters of galaxies}

\author{M. Angelinelli\inst{1,2} \fnmsep\thanks{\email{matteo.angelinelli2@unibo.it}}
          \and
          S. Ettori\inst{2,3}
          \and
          K. Dolag\inst{4,5} 
          \and
          F. Vazza\inst{1,6,7}
          \and
          A. Ragagnin\inst{1,8,9}
          }

   \institute{Dipartimento di Fisica e Astronomia, Università di Bologna, Via Gobetti 92/3, 40121 Bologna, Italy
         \and
             INAF, Osservatorio di Astrofisica e Scienza dello Spazio, via Piero Gobetti 93/3, 40121 Bologna, Italy
         \and
             INFN, Sezione di Bologna, viale Berti Pichat 6/2, 40127 Bologna, Italy
         \and
            Universitäts-Sternwarte, Fakultät für Physik, Ludwig-Maximilians-Universität München, Scheinerstr.1, 81679 München, Germany
        \and 
            Max-Planck-Institut für Astrophysik, Karl-Schwarzschild-Straße 1, 85741 Garching, Germany
         \and 
             Hamburger Sternwarte, University of Hamburg, Gojenbergsweg 112, 21029 Hamburg, Germany
         \and
            Istituto di Radio Astronomia, INAF, Via Gobetti 101, 40121 Bologna, Italy
         \and
            INAF, Osservatorio Astronomico di Trieste, via G.B. Tiepolo 11, 34143 Trieste, Italy
        \and
            IFPU, Institute for Fundamental Physics of the Universe, Via Beirut 2, 34014 Trieste, Italy
        }

   \date{Received / Accepted}


\abstract    
{

We study the redshift evolution of the baryon budget in a large set of galaxy clusters from the {\it Magneticum} suite of Smoothed Particle Hydrodynamical cosmological simulations. 
At high redshifts ($z\gtrsim1$), we obtain ``closed box'' (i.e. the baryon mass fraction $f_{\rm bar} = \Omega_{\rm bar} / \Omega_{\rm tot}$) systems independently by the mass of the systems on radii greater than $3R_{500,\mathrm c}$, whereas at lower redshifts, only the most massive halos could be considered as ``closed box''. Furthermore, in the innermost regions ($r<R_{500,\mathrm c}$), the baryon fraction shows a general decrease with the redshift and, for less massive objects, we observe a much more prominent decrease than for massive halos ($f_{\rm bar} \times \Omega_{\rm tot} / \Omega_{\rm bar} = Y_{\rm bar}$  decreases by $\sim4\%$ from $z\sim2.8$ to $z\sim0.2$ for massive systems and by $\sim15\%$ for less massive objects in the same redshift range).
The gas depletion parameter $Y_{\rm gas} = f_{\rm gas} / (\Omega_{\rm bar} / \Omega_{\rm tot})$ shows a steeper and highly scattered radial distribution in the central regions ($0.5R_{500,\mathrm c}\leq r\leq 2R_{500,\mathrm c}$) of less massive halos with respect to massive objects at all redshifts, while on larger radii ($r\geq2R_{500,\mathrm c}$) the gas fraction distributions are independent of the masses 
or the redshifts. We divide the gas content of halos into the hot and cold phases. The hot, X-rays observable, component of the gas traces well the total amount of gas at low redshifts (e.g. for $z\sim0.2$ at $R_{500,\mathrm c}$, in the most massive sub-sample --$4.6\times 10^{14}\leq M_{500,\mathrm c}/M_{\odot}\leq 7.5\times10^{14}$ / less massive sub-sample --$6.0\times 10^{14}\leq M_{500,\mathrm c}/M_{\odot}\leq 1.9\times10^{14}$-- we obtain: $Y_{\rm gas}\sim0.75/0.67$, $Y_{\rm hot}\sim0.73/0.64$, and $Y_{\rm cold}\sim 0.02 / 0.02$). 
On the other hand, at higher redshifts, the cold component provides a not negligible contribution to the total amount of baryon in our simulated systems, especially in less massive objects (e.g. for $z\sim2.8$ at $R_{500,\mathrm c}$, in the sub-sample of the most massive objects --$2.5\times 10^{13}\leq M_{500,\mathrm c}/M_{\odot}\leq 5.0\times10^{13}$ / less massive sub-sample --$5.8\times 10^{12}\leq M_{500,\mathrm c}/M_{\odot}\leq 9.7\times10^{12}$--, we measure: $Y_{\rm gas}\sim0.63 / 0.64$, $Y_{\rm hot}\sim0.50 / 0.45$, and $Y_{\rm cold}\sim0.13 / 0.18$). 
Moreover, the behaviour of the baryonic, entire gas, and hot gas phase depletion parameters as a function of radius, mass, and redshift are described by some functional forms for which we provide the best-fit parametrization. 
The evolution of metallicity and stellar mass in halos suggests that the early ($z>2$) enrichment process is dominant, while more recent star-formation processes give negligible contributions to the enrichment of the gas metallicity. In addition, Active Galactic Nuclei (AGN) have an important role in the evolution of galaxy clusters' baryon content. Thereby, we investigate possible correlations between the time evolution of AGN feedback and the depletion parameters in our numerical simulations. Interestingly, we demonstrate that the energy injected by the AGN activity shows a particularly strong positive correlation with $Y_{\rm bar}$, $Y_{\rm cold}$,$Y_{\rm star}$ and a negative one with $Y_{\rm hot}$, $Z_{\rm Tot}$. $Y_{\rm gas}$ shows the less prominent level of negative correlation, a result which is highly dependent on the mass of the halos. 
These trends are consistent with previous theoretical and numerical works, meaning that our results, combined with findings derived from current and future X-rays observations, represent possible proxies to test the AGN feedback models used in different suites of numerical simulations.

}


\maketitle

\section{Introduction} \label{sec:introduction}

The evolution of the baryon content in galaxy groups and clusters plays a key role in the understanding of the formation and growth of such systems. Indeed, it is expected that their baryon budget approaches the cosmological ratio between the cosmological baryon density $\Omega_{\rm bar}$ and the total matter density $\Omega_{\rm m}$. This condition is described as clusters of galaxies being "closed boxes" \citep{Gunn72,Bertschinger85,Voit2005}, which in turn allows neglecting the effects of feedback from galaxy formation  \citep[e.g.][]{allen11}.  
However, as we showed in our recent work \citep[][subsequently referred to as {\it PaperI}]{Angelinelli22}, non-gravitational physics related to galaxy formation significantly alters this picture, by moving a large number of baryons well beyond the virial radius of their host halos. 
Only for massive systems ($M_{vir} \geq 5\times 10^{14} h^{-1} M_{\odot}$) and at very large radii ($r \geq 6R_{500, \mathrm c}$), the baryon fraction approaches the cosmological value, verifying the condition for a “closed-box” system.

Many observational studies \citep{Sun09,Ettori15,Lovisari15,Eckert16,Nugent20} show how the baryon fraction in the central region ($<R_{500,\mathrm c}$) of galaxy groups and clusters increases with the mass of the system. Studying how it evolves with redshift is more challenging, due to the current observational limitations.  
In \citet{Gonzalez2013}, the authors analyse the baryon content in a sample of 12 galaxy clusters at $z\sim0.1$ and in the mass range between 1 and $5 \times 10^{14} \ M_{\odot}$, using XMM-Newton. They report a dependence of baryon fraction on the cluster's mass, with a slope of $\sim0.16$. Moreover, they find that less massive systems ($M_{500} \leq 2\times10^{14} \ M_{\odot}$) show a larger scatter in baryon fraction, with values which span from 60\% to 90\% of the WMAP7 \citep{2011ApJS..192...18K} cosmological expectation $\Omega_{\rm bar}/\Omega_{\rm m}$. Nevertheless, also massive systems show a depletion with respect to the cosmological expectation of $\sim18\%$. However, if the assumed cosmology is derived from Planck results \citep{2013A&A...550A.131P}, the scatter for less massive system spans from 65\% to 100\% and the depletion for massive objects decrease to 7\%, becoming consistent with the cosmological expectation, because of the systematic errors associated with the masses measurements. 
\citet{Chiu2016} study a sample of 14 galaxy clusters (with a median redshift of $z=0.9$ and masses $M_{500}=6\times10^{14} \ M_{\odot}$) selected from the South Pole Telescope (SPT) with follow-up data from XMM-Newton and Chandra telescope. They find a baryon fraction of 10.7\% with a dependency on the clusters' mass but not on the redshift. In particular, the authors suggest that the slope of the $f_{\rm bar}-M_{500}$ relation is $\sim0.22$, while the uncertainties on the mass estimations introduce an uncertainty in the redshift trend parameter which is larger than the statistical uncertainty, making impossible any clear evidence of a redshift dependency. Given the relations, the authors conclude that a simple hierarchical structure formation merger model is not sufficient to completely describe the accretion of galaxy clusters or groups. Significant accretion of galaxies and intracluster medium (ICM) from the field, combing with the loss of stellar mass from galaxies through stripping, are needed to completely explain the observational finding they discussed. 
More recently, \citet{Akino22} study a sample of 136 galaxy clusters and groups with $M_{500}$ masses between $10^{13}$ up to $10^{15} \ M_{\odot}$ and a redshift range which spans from 0 to 1. They perform a joint analysis using HSC-SSP weak-lensing mass measurements, XXL X-ray gas mass measurements, and HSC
and Sloan Digital Sky Survey multiband photometry. They find that the baryon fraction systems mass relation shows steeping of the slope moving from group regime to cluster one. Moreover, they find that the baryon fraction is $\sim50\%$ for $\sim10^{13}\ M_{\odot}$, $\sim60\%$ for $\sim10^{14}\ M_{\odot}$ and $\sim100\%$ for $\sim10^{15}\ M_{\odot}$ systems with respect to the cosmological expectation $\Omega_{\rm bar}/\Omega_{\rm m}$, assumed from \citet{Planck2020}. Even if the relation between the baryon fraction and the systems' mass is observed, for the baryon fraction-redshift it is not possible to obtain strong constraints because of uncertainties in the mass estimations.

Using a semi-analytic model that connects the ``universal'' behaviour of the thermodynamic profiles with the integrated properties of the ICM by modelling the departure from self-similarity also including a dependency of the gas mass fraction within $R_{500}$ on the gas temperature and redshift, \cite{ettori22} constrain the former to be about $T^{0.4}$ and the latter in being almost negligible through the calibrations with a collection of recent published scaling laws.

\begin{figure*}
\includegraphics[width=0.49\textwidth]{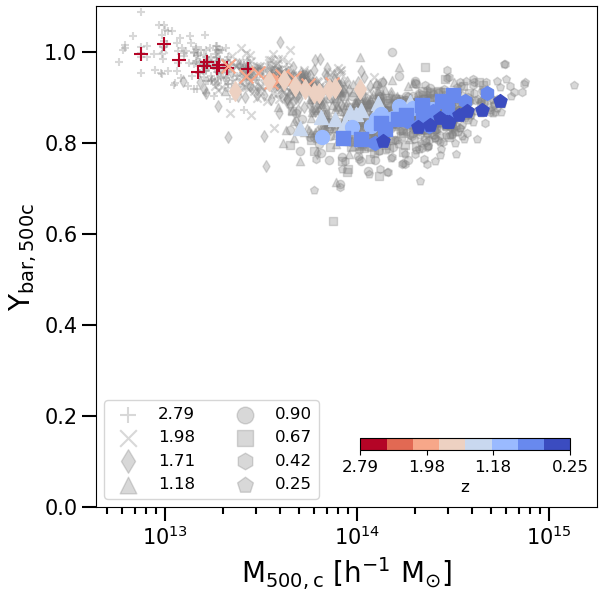}
\includegraphics[width=0.49\textwidth]{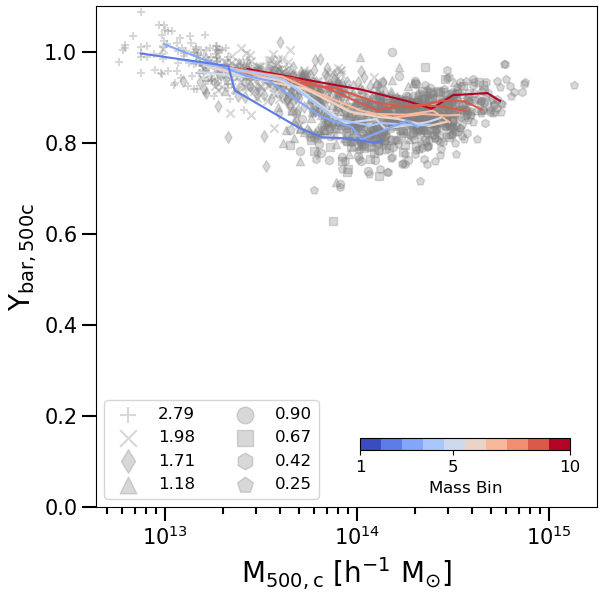}
\caption{Baryon depletion parameter inside $R_{500,\mathrm c}$. The grey dots represent single galaxy clusters, identified by different markers, accordingly to the legend in the bottom left corner. (Left) The coloured dots represent the median values computed in each of the mass bins of a single snapshot. The colour coding is given by the snapshot's redshift, following the colourbar in the bottom right corner. (Right) The coloured lines represent the redshift evolution of each mass bin. The colour coding is given by the mass bin, following the colourbar in the bottom right corner.}
\label{fig:ybar_r500}
\end{figure*}

By connecting the co-evolution of galaxies and AGN in groups and clusters of galaxies and the induced circulation of baryons, many numerical works have explored the evolution of baryon fraction across cosmic time. 
\citet{Duffy2010} used a sample of galaxy clusters extracted from the OverWhelmingly Large Simulations project \citep{Schaye2010}. 
They find that simulations with strong feedback (both from AGN or Supernovae) decrease the baryon fraction on galaxy-scale haloes by a factor of 2 or 3. On groups and cluster scales, only simulations that include appropriate levels of AGN feedback can reduce the observed baryon fraction, at least within a factor $\sim 2$. 
Simulations that include inefficient cooling and stellar feedback, as well as the ones with strong feedback models, well reproduce the stellar fraction for massive objects. On the other hand, only the simulations with strong AGN feedback reproduce the observed star formation efficiencies. 
\citet{Planelles13} use a set of simulations using the TreePM–SPH GADGET-3 code, including a combination of stellar and AGN feedback and non-radiative effects. They find that for non-radiative and stellar-only feedback runs, the baryon fraction with $R_{500}$ does not show any strong dependencies by the mass of the central clusters and it deviates from the cosmological expectation at large at $\sim10\%$. On the other hand, AGN feedback is responsible for the depletion of baryon content in galaxy group mass regime, and only for massive systems, the cosmic value is reached. Moreover, they study possible dependencies of baryon fraction with radius from cluster's centre, system's mass, and redshift. They do not find any particular trend and they suggest that further improvements could be related to the extension of the simulations with other feedback models. 
\citet{Henden2020} analyse the baryon content in the Feedback Acting on Baryons
in Large-scale Environments (FABLE) simulations. These simulations are performed using the AREPO code \citep{Springel2010}. The prescription for stellar and AGN feedback are revisited versions of the models \citep{Henden2018} adopted in the Illustis simulations \citep{Vogelsberger2014}. They find a good agreement between their findings and the observational proxies given by the X-ray observations. This implies that, when weak lensing measurements are considered and the hydrostatic mass bias is taken into account, the systems they analysed result too gas rich,  meaning that the models must be revisited in order to reproduce the most accurate observational constraints 
Moreover, their findings suggest that there is a different evolution with cosmic time in systems with different masses. Indeed, for massive systems ($M_{500}>3\times10^{14} \ M_{\odot}$) the total gas and stellar mass are approximately independent of redshift at $z\leq1$. Otherwise, less massive systems show a significant redshift evolution. The authors conclude that this is important for understanding the different growth of massive galaxy clusters and smaller systems. For the former is expected that they accumulated mass accreting low mass systems, while these later seem to show little redshift evolution themselves.
\citet{Davies2020} compare results from EAGLE \citep{Schaye2015,Crain2015} and Illustris-TNG \citep{Pillepich2018,Nelson2018,Springel2018} simulations. Even if these simulations share aims and scope, they are very different in the recipes adopted for hydrodynamics solvers and the solutions of the physical processes included, mainly for the feedback one. 
In their work, the authors focus on the properties of the circumgalactic medium (CGM) and the quenching and morphological evolution of central galaxies. They find that in both EAGLE and Illustris-TNG simulations, the influence of halo properties on central galaxies is mainly driven by the expulsion of CGM. Moreover, feedback is also responsible for the heating of the remaining CGM, which contributes to the growth of the cooling time and inhibits the accretion of gas. The results are similar in both the suits used, but there are also some differences which will be in principle tested from an observational point of view. Indeed, studying the scaling relations between the column density of CGM \textsc{Ovi} absorbers and the specific star formation rate of central galaxies at fixed halo mass, or between the CGM mass fraction of haloes and the accretion rate of their central Black Hole (BH), it is possible to disentangle between the different models adopted in the different simulations, which predict different scenario for these relations. The authors conclude that, even if some differences are observed between these simulations, the role of the AGN feedback on the CGM and central galaxies is dominant in the entire cosmic evolution of such systems.  
Recently, \citet{Robson2023} study a sample of simulated galaxy clusters and groups, with masses $M_{500}$ from $10^{12.3}M_{\odot}$ to $10^{15} M_{\odot}$, extracted from the SIMBA simulations \citep{Dave2019}. They analyse the evolution of the X- ray scaling relations and X-ray profiles from $z=3$ to $z=0$. Moreover, they study the impact of different feedback models in comparison with the self-similar evolution. They find that halos show a consistent slope with the self-similar one for $z>1.5$, while at lower redshifts the number of groups that deviate from self-similarity increase. Regarding the relation between gas fraction and halo mass, they observe a drop and increasing in the scatter with redshifts $z<1.5$, especially for halos with $M_{500}<10^{13.5}M_{\odot}$. Comparing simulations which include or exclude different feedback models, they observe that the only the AGN feedback is able to highly influence the scaling relations they analysed. In particular, they find that for halos with $M_{500}<10^{13.5}M_{\odot}$ the gas fraction is lowered by the AGN feedback, meaning reduction of X-rays luminosity and temperature of these systems. On the other hand, the gas metallicity seems to be the only parameter that is more influenced by stellar feedback with respect to AGN one. \citet{Robson2023} highlight that their analysis wants to address the connection between galaxy quenching and X-ray properties across cosmic time and their results could be useful as basis for comparison with other physical models and future observations.  


In this work, we perform a detailed analysis of the baryon content of galaxy clusters and its redshift evolution in a sample of halos simulated in the \textit{Magneticum}\footnote{http://www.magneticum.org} suite, understating the role of gas, and its different phases, and the stellar component, as well as the correlations between AGN feedback energy and the possible traces of these interactions in the evolution of the baryonic distribution. The paper is structured as follows: in Sect.~\ref{sec:methods}, we briefly describe the \textit{Magneticum} simulations and how we select our sample; we present the results of our analysis and the comparison with recent observational in Sect.~\ref{sec:results}; we discuss our main findings and the correlations between AGN feedback and baryons evolution in Sect.~\ref{sec:discussion}, while in Sect.~\ref{sec:conclusions}, we summarise our results and possible future extensions of this work. 

\begin{figure*}
\includegraphics[width=0.49\textwidth]{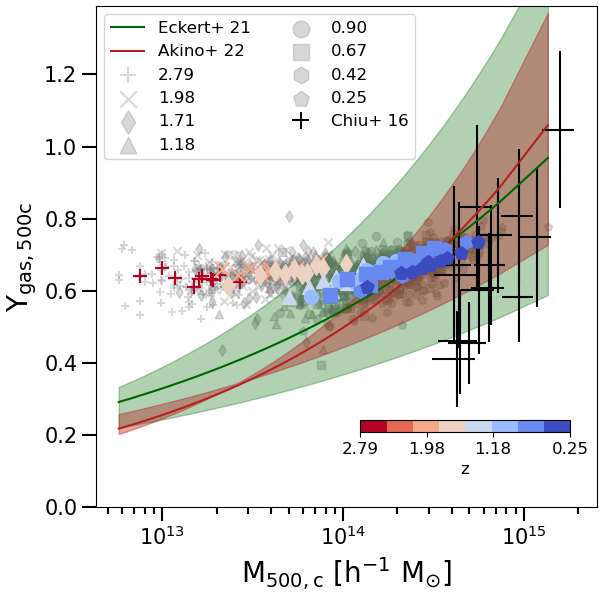}
\includegraphics[width=0.49\textwidth]{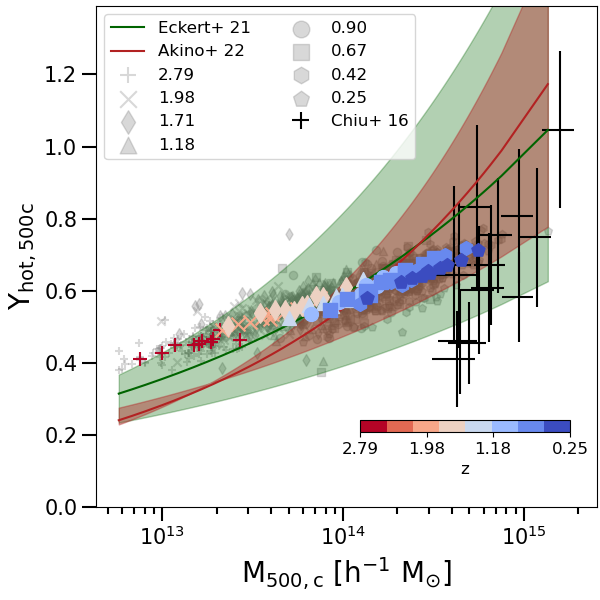}
\caption{Gas depletion parameter (Left) and hot gas phase depletion parameter (right) inside $R_{500,\mathrm c}$. The grey dots represent single galaxy clusters, identified by different markers, accordingly to the legend in the bottom right corner. The coloured dots represent the median values computed in each of the mass bins of a single snapshot. The colour coding is given by the snapshot's redshift, following the colourbar in the top left corner. The solid lines (and related shadow regions) represent the fit proposed by \citet{Eckert21} ($f_{\rm gas,500} = 0.079^{+0.026}_{-0.025} ( M_{500} / 10^{14} M_{\odot} )^{0.22^{+0.06}_{-0.04}}$, shown in green) and \citet{Akino22} ($ln (M_{\rm gas}/10^{12} M_{\odot} ) = 1.95^{+0.08}_{-0.08} + 1.29^{+0.16}_{-0.10} ln(M_{500}/10^{14} M_{\odot})$, shown in red). The black crosses represent the $f_{\rm gas}$ estimates in \citet{Chiu2016}.}
\label{fig:ygasyhot_r500}
\end{figure*}

\section{The {\it Magneticum} cosmological simulations} \label{sec:methods}

We selected a sub-sample of galaxy clusters from  \textit{Box2b/hr} of \textit{Magneticum} simulations at eight different snapshots, corresponding to redshifts 2.79,1.98,1.71,1.18,0.90,0.67,0.42 and 0.25. The high-resolution run of \textit{Box2b} includes a total of $2\cdot2880^3$ particles in a volume of $(640 \ h^{-1}\rm cMpc)^3$. The particles masses are $6.9\cdot10^8 \ h^{-1}\rm M_{\odot}$ and $1.4\cdot10^8 \ h^{-1}\rm M_{\odot}$, respectively for dark matter and gas component and the stellar particles have softening of $\epsilon=2 \ h^{-1}\rm ckpc$. The cosmology adopted for these simulations is the WMAP7 from \citet{2011ApJS..192...18K}, with a total matter density of $\Omega_{\rm m}=0.272$, of which 16.8$\%$ of baryons, the cosmological constant $\rm \Lambda_{0}=0.728$, the Hubble constant $\rm H_{0}=70.4 \ \rm km/s/Mpc$, the index of the primordial power spectrum $\rm n=0.963$ and the overall normalisation of the power spectrum $\sigma_{8}=0.809$. 
The more relevant physical mechanisms included in \textit{Magneticum} are:
cooling, star formation and winds with velocities of $\rm 350 \ km/s$ \citep{2002MNRAS.333..649S}; 
tracing explicitly metal species (namely, C, Ca, O, N, Ne, Mg, S, Si, and Fe) and following in detail the stellar population and chemical enrichment by SN-Ia, SN-II, AGB \citep{2003MNRAS.342.1025T,2007MNRAS.382.1050T} and cooling tables from \citet{2009MNRAS.399..574W};
 black holes and associated AGN feedback \citep{2005MNRAS.361..776S} with various improvements \citep{2010MNRAS.401.1670F,2014MNRAS.442.2304H} for the treatment of the black hole sink particles and the different feedback modes;
 isotropic thermal conduction of 1/20 of standard Spitzer value \citep{2004ApJ...606L..97D};
 low viscosity scheme to track turbulence \citep{2005MNRAS.364..753D,2016MNRAS.455.2110B};
 higher order SPH kernels \citep{2012MNRAS.425.1068D};
 passive magnetic fields \citep{2009MNRAS.398.1678D}.
Halos are identified using \textsc{subfind} \citep{2001MNRAS.328..726S, 2009MNRAS.399..497D}, where the centre of a halo is defined as the position of the particle with the minimum of the gravitational potential. The virial mass, $M_{vir}$ is defined through the spherical overdensity as predicted by the generalised spherical top-hat collapse model \citep{1996MNRAS.282..263E} and, in particular, it is referred to $R_{vir}$, whose overdensity to the critical density follows Eq.~6 of \citet{Bryan98}, which correspond to $\approx117$ for the given redshift and cosmology. The radii $R_{200,\mathrm m}$ and $R_{500,\mathrm c}$ are defined as a spherical-overdensity of 200 (respectively 500) to the mean (respectively critical) density in the chosen cosmology.  

As shown in previous studies, the galaxy physics implemented in the \textit{Magneticum} simulations leads to an overall successful reproduction of the  basic galaxy properties, like the stellar mass-function \citep{2017ARA&A..55...59N,2022arXiv220109068L}, the environmental impact of galaxy clusters on galaxy properties \citep{2019MNRAS.488.5370L} and the appearance of post-starburst galaxies \citep{2021MNRAS.506.4516L} as well as the associated AGN population at various redshifts \citep{2014MNRAS.442.2304H,2016MNRAS.458.1013S,2018MNRAS.481.2213B}. At cluster scales, the \textit{Magneticum} simulations have demonstrated to reproduce the observable X-ray luminosity-relation \citep{2013MNRAS.428.1395B}, the pressure profile of the ICM \citep{2017MNRAS.469.3069G} and the chemical composition \citep{2017Galax...5...35D,2018SSRv..214..123B} of the ICM, the high concentration observed in fossil groups \citep{2019MNRAS.486.4001R},  as well as the gas properties in between galaxy clusters \citep{Biffi22}. On larger scales, the \textit{Magneticum} simulations demonstrated to reproduce the observed SZ-Power spectrum \cite{2016MNRAS.463.1797D} as well as the observed thermal history of the Universe \citep{2021PhRvD.104h3538Y}.    

In each selected snapshot, we selected the 150 most massive galaxy halos. The final sample (combining all the different snapshots) is composed of 1200 galaxy clusters, described by a $M_{500, \mathrm c}$ mass range between $\sim10^{13} \ h^{-1}M_{\odot}$ and $\sim10^{15} \ h^{-1}M_{\odot}$ (see Tab.~\ref{tab:mass_redshift} for details on the mass range). Moreover, for each snapshot, we divided the sample into 10 equal bins into the logarithmic space, so that each bin contains 15 objects.
As described in our previous work {\it PaperI}, we extend our radial analysis up to $10R_{500,\mathrm c}$. Firstly, we consider the accretion shocks position as the location of the peak of the entropy profile \citep[as proposed in][]{2011MNRAS.418..960V} and we find that in our systems it is closer to $\sim 6R_{500,\mathrm c}$. The accretion shocks position is often used as the boundary of galaxy clusters \citep[see][for details on the accretion shocks definitions and expected locations]{Zhang20,Aung21} and extending our analysis in regions external to the accretion shocks ensures that we are mapping the entire volume of a given halo and allows us to characterise its baryon and gas mass fraction. 

\begin{table}[]
\begin{tabular}{c|cc|cc|}
&\multicolumn{2}{c|}{$M_{500,\mathrm c} \ [h^{-1} M_{\odot}]$}&\multicolumn{2}{c|}{$M_{vir} \ [h^{-1} M_{\odot}]$} \\ \hline
z&Min&Max&Min&Max \\ \hline \hline
2.79&$5.8\times10^{12}$&$6.3\times10^{13}$&$9.2\times10^{12}$&$8.0\times10^{13}$ \\
1.98&$9.8\times10^{12}$&$1.2\times10^{14}$&$1.8\times10^{13}$&$1.8\times10^{14}$ \\
1.71&$1.2\times10^{13}$&$1.5\times10^{14}$&$3.0\times10^{13}$&$2.4\times10^{14}$ \\
1.18&$2.1\times10^{13}$&$3.2\times10^{14}$&$4.6\times10^{13}$&$4.5\times10^{14}$ \\
0.90&$4.5\times10^{13}$&$3.5\times10^{14}$&$9.7\times10^{13}$&$5.6\times10^{14}$ \\
0.67&$4.5\times10^{13}$&$4.6\times10^{14}$&$1.2\times10^{14}$&$7.0\times10^{14}$ \\
0.42&$6.1\times10^{13}$&$7.4\times10^{14}$&$1.5\times10^{14}$&$1.3\times10^{15}$ \\
0.25&$6.0\times10^{13}$&$1.4\times10^{15}$&$1.5\times10^{14}$&$2.0\times10^{15}$ \\ \hline
\multicolumn{5}{c}{} \\
\end{tabular}
\caption{Minimum and maximum halos masses ($M_{500,\mathrm c}$ and $M_{vir}$) for each selected redshift.}
\label{tab:mass_redshift}
\end{table}

The dynamics of accreting gas, which mostly gets shock heated during its first infall, is different from that of the collisionless dark matter. Thus, while the radius of accretion shocks defines the spatial extent of the gas in the DM halos, a different boundary must be defined using DM particles only \citep[e.g.][]{walker19}.
One example is the splashback radius $R_{sp}$ \citep{Adhikari14}, which represents the apocenters (farthest point of the particle orbit with respect to the halo potential minimum) of infalling Dark Matter through the pericenter.
Many different works\footnote{see http://www.benediktdiemer.com/research/splashback/ for a complete bibliography about the splashback radius} have  demonstrated that $R_{sp}$ is $\sim2.5R_{200, \rm c}$. Being $R_{200, \rm c}\sim1.6R_{500,\rm c}$, $R_{sp}$ is $\sim4R_{500, \rm c}$, which is closer to halo centre compared to the radius of the accretion shock. This means that the assumption of the accretion shock radius as a boundary of the halo, combined with an extension of the analysis on radii larger than the accretion shock position, ensures we consider in our work all the baryons that are enclosed in the simulated halos.

\section{The cosmic depletion parameters} \label{sec:results}

We are interested in the cosmological evolution of different matter components within simulated groups and clusters and similar to \textit{PaperI}, we want to focus on 
 volume-integrated quantities as baryon, gas and star fraction:
\begin{align}
  f_{\rm bar}(<r) &= (m_{\rm gas}(<r)+m_{\rm star}(<r)+m_{\rm BH}(<r))/m_{\rm tot}(<r) \\
  f_{\rm gas}(<r) &= m_{\rm gas}(<r)/m_{\rm tot}(<r) \\
  f_{\rm star}(<r) &= m_{\rm star}(<r)/m_{\rm tot}(<r)
\end{align}
where $r$ is the radial distance from the cluster centre. The different $m_{\rm i}(<r)$ are referred to as different particles type (gas, stars, or black holes), while $m_{\rm tot}(<r)$ is the sum of the previous masses and the dark matter up to the radial shell $r$. 
We divide the gas component of our systems into two different phases, $f_{\rm hot}$ which considers the gas particles with a temperature greater than 0.1 keV, and $f_{\rm cold}$ for the remaining ones:
\begin{align}
    f_{\rm hot}(<r) &= m_{\rm hot}(<r)/m_{\rm tot}(<r) \\
    f_{\rm cold}(<r) &= m_{\rm cold}(<r)/m_{\rm tot}(<r).
\end{align}
Finally, we refer to the cosmic depletion parameter $Y$, defined as
\begin{equation}
    Y(<r) = f(<r)/(\Omega_{\rm bar}/\Omega_{\rm m})
\end{equation}
where $f(<r)$ could assume any definition given above and $\Omega_{\rm bar}/\Omega_{\rm m}=0.168$, the cosmological value of baryon over total matter adopted for \textit{Magneticum} simulations.
The 0.1 keV ($\sim10^6 K$) threshold is used as a conservative lower limit on the gas temperature to consider the gas that can be detected by X-rays observatories \citep[see][for details on the comparison between simulations and X-rays observations]{mazzotta04,Rasia2014,Biffi22}.

\begin{figure}
\includegraphics[width=0.49\textwidth]{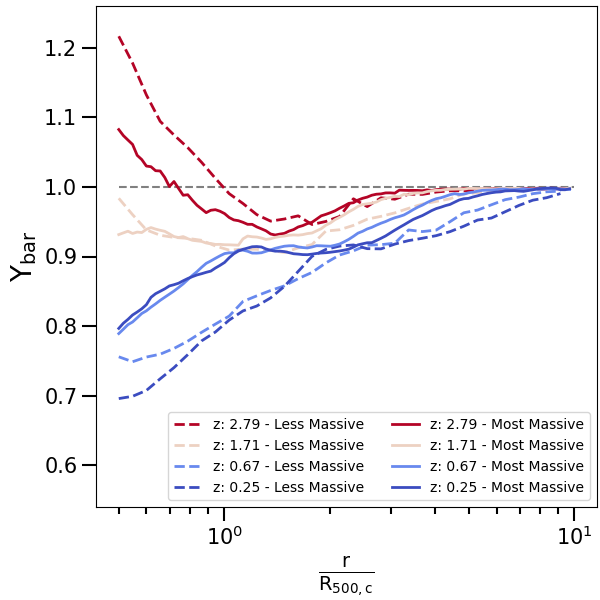}
\caption{Radial profiles of baryon depletion parameter, from $0.5R_{500, \mathrm c}$ up to $10R_{500, \mathrm c}$, for the less massive bin (dashed lines) and most massive one (solid lines). The lines represent the median profiles at four different redshifts, accordingly to the colours in the bottom right corner
(see Tab.~\ref{tab:fixradii_redshift} for the definition of the mass ranges).}
\label{fig:ybar_radial}
\end{figure}

\begin{figure}
\includegraphics[width=0.49\textwidth]{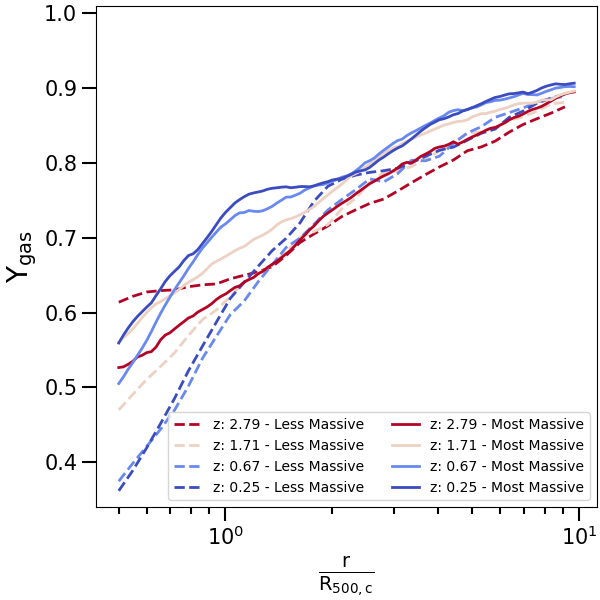}
\caption{Radial profiles of gas depletion parameter, from $0.5R_{500, \mathrm c}$ up to $10R_{500, \mathrm c}$, for the less massive bin (dashed lines) and most massive one (solid lines). The lines represent the median profiles at four different redshifts, accordingly to the colours in the bottom right corner (see Tab.~\ref{tab:fixradii_redshift} for the definition of the mass ranges).}
\label{fig:ygas_radial}
\end{figure}

\subsection{Depletion parameters within $R_{500,\mathrm c}$} \label{sec:resultsr500}

Firstly, we focus on the innermost regions of galaxy clusters. In Fig.~\ref{fig:ybar_r500} we show the baryon depletion parameter inside $R_{500,\mathrm c}$  as a function of the host cluster's mass. From the left plot of Fig.~\ref{fig:ybar_r500}, we observe a general decrease of the baryon depletion parameter across the cosmic time. From the right panel, we follow the redshift evolution of the baryon depletion parameter of a single mass bin. Massive objects show a flatter behaviour than less massive systems, and for the latter, the baryon depletion parameter decreases by $\sim15\%$ from $z=2.79$ to $z=0.25$ (see Tab.~\ref{tab:fixradii_redshift}).

We compare our findings with the observational constraints within $R_{500,\mathrm c}$, from recent work by \citet{Eckert21} and \citet{Akino22}, where different best fits were proposed to  describe the  gas fraction as a function of the host cluster's mass. 
In Fig.~\ref{fig:ygasyhot_r500}, we show these best fits against our findings on gas and hot gas phase depletion parameters, also including the results proposed by \citet{Chiu2016} and already introduced in Sect.~\ref{sec:introduction}.
As in {\it PaperI}, our results are able to correctly reproduce the observational findings, and for low  redshift ($z<1.2$) halos show an increase of the gas fraction with the cluster's mass. 
On the other hand, in high redshift ($z>1.2$) halos the gas fraction appears to be independent of the mass of the central cluster, with values $Y_{\rm gas}\sim0.65$. In the right plot of Fig.~\ref{fig:ygasyhot_r500}, we show the hot gas phase depletion parameter inside $R_{500,\mathrm c}$ as a function of cluster mass. Comparing the left and right plots of Fig.~\ref{fig:ygasyhot_r500}, we notice that, for low redshift systems, the hot gas phase is able to completely recover the total gas depletion fraction. On the other hand, for high redshift systems, the hot component is always a fraction of the total gas amount of galaxy clusters. This implies that for high redshift systems the cold gas component is far from negligible, not even from the most massive halos. This suggests that in forming systems closer to their formation time the virialization process is far from complete, and large fractions of the gas mass are still cold; this also suggests that our earlier findings in 
 {\it PaperI} become increasingly less accurate moving to higher redshifts.
 
In this respect, the differences between the hot gas phase component and the total amount of gas embedded in high redshift galaxy clusters have particular importance in the study of proto-galaxy clusters. Indeed, these objects, characterised by relatively low masses and high redshifts, seem to show the highest displacement between the real amount of gas and the one which is recovered by X-rays observations. Further investigations are needed to completely assess these differences and understand how to take into account them in the computation of real proto-cluster masses \citep[see][for a review on proto-clusters]{Overzier2016}.  

\begin{figure}
\includegraphics[width=0.49\textwidth]{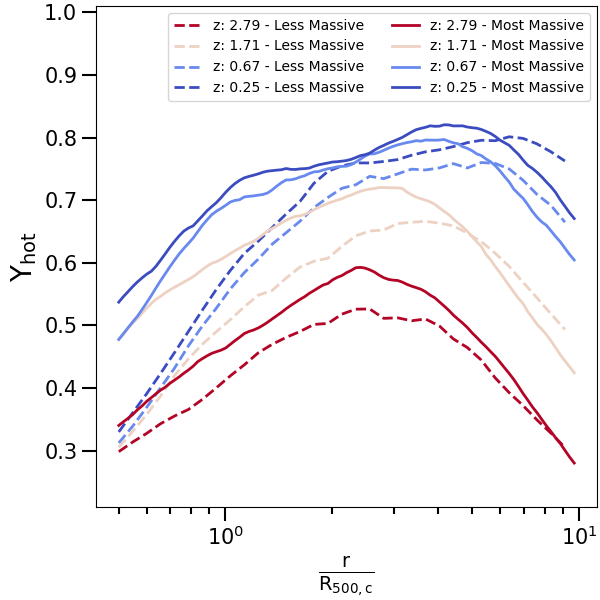}
\caption{Radial profiles of the depletion parameter for the hot gas phase, from $0.5R_{500, \mathrm c}$ up to $10R_{500, \mathrm c}$, for the less massive bin (dashed lines) and most massive one (solid lines). The lines represent the median profiles at four different redshifts, accordingly to the colours in the top right corner (see Tab.~\ref{tab:fixradii_redshift} for the definition of the mass ranges).}
\label{fig:yhot_radial}
\end{figure}

\subsection{Depletion parameters up to $10R_{500, \mathrm c}$} \label{sec:resultsradial}

In Fig.~\ref{fig:ybar_radial}, we show the median distributions of the baryon depletion parameters as a function of the radius (between $0.5R_{500, \mathrm c}$ and $10R_{500, \mathrm c}$) for the less massive and most massive mass bins, computed in each redshift. In Fig.~\ref{fig:ybar_radial}, we identify a redshift evolution of the profiles. Indeed, in the innermost regions ($r<R_{500, \mathrm c}$) we observe a decreasing of $\sim0.40\%$ moving from $z=2.79$ to $z=0.25$. In regions far from the cluster's centre ($r<5R_{500, \mathrm c}$), the differences between low and high redshift systems are less than $10\%$ from less massive objects and less than $5\%$ for massive ones. At low redshifts, the "closed-box" assumption remains true only for massive objects and on radii greater than $5R_{500, \mathrm c}$ (see also {\it PaperI}). At high redshifts, the same condition is reached independently by the halo mass and on radii closer to $3R_{500, \mathrm c}$. 

In Fig.~\ref{fig:ygas_radial}, we present the median gas depletion parameters as a function of the radius, for the less massive and most massive mass bins, computed in each redshift investigated. At high redshifts, the differences between low and high mass objects are less than $10\%$, while, at low redshifts, the same differences are larger than $20\%$. Fig.~\ref{fig:ygas_radial} shows an increase in the gas content with the radius. This increase is steeper for less massive objects at low redshift, whereas for massive systems the trend is rather redshift-independent. In each of the analysis cases, the gas depletion parameter approaches values between $85\%$ and $90\%$ at $10R_{500, \mathrm c}$, independent of the mass or the redshift.

In Fig.~\ref{fig:yhot_radial} we show the median hot gas phase depletion parameters as a function of the distance from the cluster's centre, for the less massive and most massive mass bins, computed in each redshift.  Differently from the profiles of baryon and gas depletion parameters presented in Fig.~\ref{fig:ybar_radial} and Fig.~\ref{fig:ygas_radial}, here the profiles show a marked peak and the following drop. This trend is already discussed in {\it PaperI}, where we surmised that the position of the peak is closer to the position of the accretion shock. Interestingly, here we can further observe a shift to the outer regions of the peak with the decrease of redshift. This is compatible with an increase in the halo volume with cosmic time, marked by the expansion of the outer accretion regions. We also notice that most massive objects have a higher contribution of hot gas at every redshift. Moreover, comparing Fig.~\ref{fig:ygas_radial} and Fig.~\ref{fig:yhot_radial}, we note that for low redshift systems the total amount of gas within the cluster's volume is quite perfectly traced by the hot component, while for high redshift objects the hot gas phase is always less than $75\%$ of the total gas. Therefore, in high redshift galaxy clusters, the hot and X-rays observable part of the gas represents only a fraction of the total gas mass, making it indispensable to correct the derived mass in order to make any accurate cosmological use of it. 

\subsection{Gas metallicity and stellar component} \label{sec:resultsmet}

\begin{figure}
\includegraphics[width=0.49\textwidth]{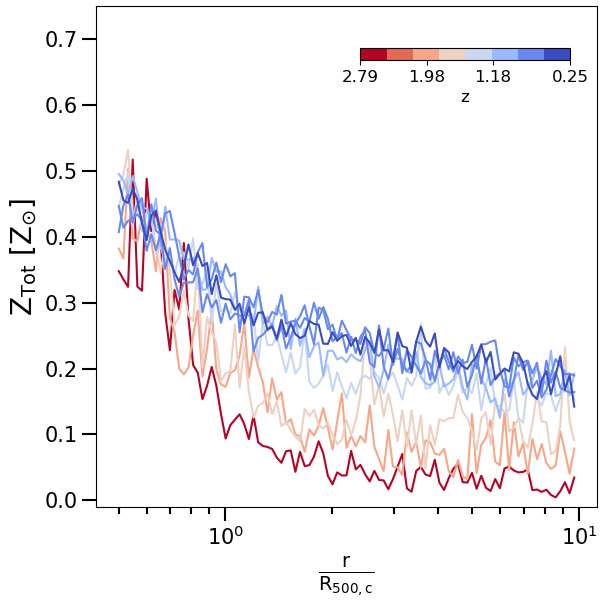}
\caption{Radial profiles of gas metallicity (in Solar units), from $0.5R_{500, \mathrm c}$ up to $10R_{500, \mathrm c}$, for the less massive bin (dashed lines) and most massive one (solid lines). The lines represent the median profiles at four different redshifts, accordingly the colours in the top right corner (see Tab.~\ref{tab:fixradii_redshift} for the definition of the mass ranges).}
\label{fig:ztot_radial}
\end{figure}

The injection and evolution of metals by SN-Ia, SN-II, and AGB stars in \textit{Magneticum} simulations are modeled following \citet{2003MNRAS.342.1025T,2007MNRAS.382.1050T}.
As already done in {\it PaperI}, we consider the total metallicity as the sum of the elements heavier than helium relative to the hydrogen mass.
The total metallicity at each radial shell $r$, $Z_{\rm tot}(r)$, is the mass-weighted sum of the metallicity of the gas particles $i$ with mass $m_{\rm gas, i}$ which belong to the radial shell $r$:
\begin{equation}
    Z_{\rm tot}(r) = \frac{\sum_{\rm i} Z_{\rm tot,i} \cdot m_{\rm gas,i}}{\sum_{\rm i} m_{\rm gas,i}}.
\end{equation}
The radial shells are defined to include a fixed number of 250 particles, to allow a significant statistical analysis of each of them. We normalise these values of metallicity to the Solar values proposed by \citet{Asplund09}: $Z_{\odot}=0.0142$.
In Fig.~\ref{fig:ztot_radial}, we show the median distributions of the total metallicity as a function of the radius for the less massive and most massive mass bins, computed in each redshift. Here we note that the profiles are highly scattered, as already discussed and justified in {\it PaperI}. Moreover, although no strong mass dependencies are observed, we notice a clear evolution of total metallicity across time. Indeed, independently of the mass bin analysed, the values of metallicity increase towards lower redshifts.  In the external part of the galaxy clusters, on radii larger than $2\div3R_{500, \mathrm c}$, we observe a general flatting of the profiles. This trend is slightly prominent in high redshift systems. 

\begin{figure}
\includegraphics[width=0.49\textwidth]{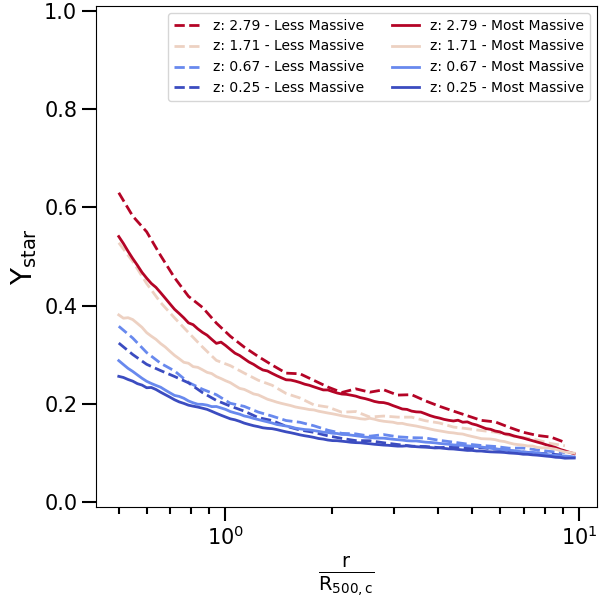}
\caption{Radial profiles of stellar depletion parameter, from $0.5R_{500, \mathrm c}$ up to $10R_{500, \mathrm c}$, for the less massive bin (dashed lines) and most massive one (solid lines). The lines represent the median profiles at four different redshifts, accordingly to the colours in the top right corner (see Tab.~\ref{tab:fixradii_redshift} for the definition of the mass ranges).}
\label{fig:ystar_radial}
\end{figure}

In Fig.~\ref{fig:ystar_radial}, we give the median distribution of stellar depletion parameter, as a function of radius, for less massive and most massive objects, computed in each redshift. Differently from the case of baryons, gas, and hot gas phases, here we observe a gradual decrease of the stellar depletion parameter with increasing distance from the cluster's center. Instead,  similar to the metallicity profiles, also for the stellar depletion factor we do not observe any mass-associated trend. However, we observe a clear trend: especially in the innermost regions, the values of the stellar depletion parameter decrease with the redshift, 
while in the outskirts of halos the differences between different redshifts are of the order of a few percent. Halos in all mass bins approach a 
stellar depletion parameter $Y_{\rm star}\sim0.1$ a the boundary of the analysed volumes ($\sim10R_{500, \mathrm c}$). 

\section{Discussion} \label{sec:discussion}

Recently, many different works have explored the evolution of baryons in simulated halos \citep{Castro2021,Ragagnin2022,Ayromlou22,Robson2023}. In our work, we analyse a sample of galaxy clusters extracted from \textit{Magneticum} simulations. Compared to other works, the simulated volume of \textit{Box2b} is large enough to allow a selection of very massive objects ($M_{500,\rm c}>10^{15} M_{\odot}/h$ at $z=0$). Moreover, as also discussed by \citet{Robson2023} (see their Fig.10 and related discussion), the effects of different feedback phenomena highly influence the evolution of baryons in the simulated halos. From the comparison between \textit{Magneticum} simulations and other suites of numerical simulations, we can assess the impact of different feedback models in the evolution of baryons in simulated in halos. We select for each simulated snapshot the 150 most massive halos (see Tab.~\ref{tab:mass_redshift} for details on the mass range). This selection minimises the effect of mergers and phenomena that occur in the history of single objects. Indeed, in the most massive objects gravity can compensate for the effect of AGN feedback, and the evolution we recover is less affected by internal feedback phenomena. On the other hand, the other possible approach is to trace the time evolution of a sample of massive objects at $z=0$. In this case, the results are probably affected by the history of the single objects, and the evolution of baryon may be influenced by mergers or peculiar phenomena that occur in specific objects, making the results less statistically robust. Moreover, this approach does not ensure that at high redshifts the sample still remains mass-complete. Indeed, since the mass growth of halos is determined by merger phenomena peculiar to individual objects, halos that at high-$z$ belong to the most massive mass bin, may in their history not grow to sufficient masses to still belong to the most massive sub-sample at $z=0$. Furthermore, due to observational limitations, the galaxy cluster observed samples are composed of very massive objects, especially for high-$z$ observations. The selection we decided to adopt in our work is therefore closer to what can be reproduced to date with X-rays observations. \citet{Robson2023} also highlight the major role of the feedback model in the estimations of proprieties of galaxy clusters and groups derived from X-rays observations. It is therefore essential to build simulated samples as similar as possible to those that can be obtained from X-rays observations. This will allow a comparison between different feedback models used in simulations that can give results comparable with observations and thus enable a better understanding of the real effect of feedback phenomena on the evolution of baryons in halos. We defer this to future work.

\citet{Castro2021} investigate the role of AGN feedback in the halo accretion history in the \textit{Magneticum} simulations. The energy introduced in the surrounding environment by the AGN is proportional to the mass accretion rate of the black hole $\rm \dot{M}_{BH}$ \citep{sp05,2014MNRAS.442.2304H}. The authors conclude that the AGN feedback has a nearly time-universal behaviour. They find that the peak of the AGN feedback occurs at a slightly higher redshift than the baryon fraction peak, then they observe a quick decaying around $z=1$, followed by a slow decaying phase at lower redshifts. Moreover, they note a rather universal trend response to the AGN activity. The variation of the halo mass shows a significant and negative correlation with the intensity of AGN feedback when halo progenitors reach $\sim30\div50\%$ of their final mass. \citet{Castro2021} conclude that the decrease of halo mass observed in simulations is driven by the action of AGN feedback in a relatively early phase of the halo assembly when the shallower galaxy cluster's potential well can better react to the displacement of gas heated by feedback. 

Recently, \citet{Ragagnin2022} study the ejection of gas from the halo, due to AGN feedback in \textit{Magneticum} simulations. They find that the gas fraction in galaxy clusters with a redshift formation greater than 1 is lower than the one observed for systems with lower redshift formation. This difference is associated with the amount of gas present at the epoch of formation and later ejected by the AGN activity. Indeed, when the amount of ejected gas is taken into account, the distributions of gas fraction recovered are independent of the formation redshift of galaxy clusters. 

\begin{figure*}
\includegraphics[width=0.49\textwidth]{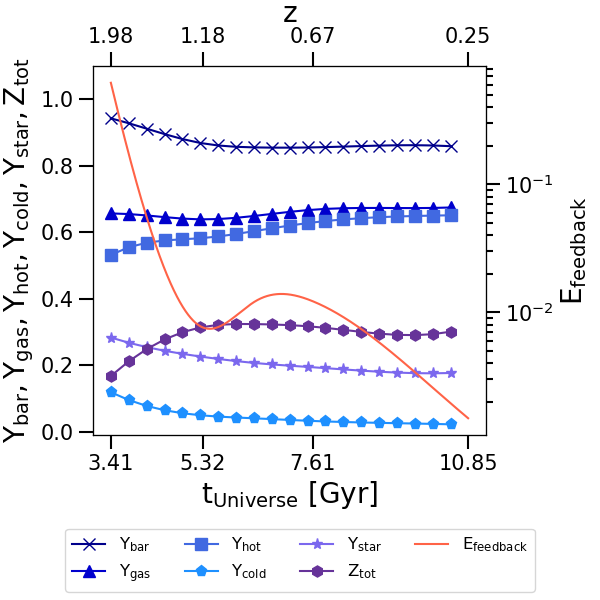}
\includegraphics[width=0.49\textwidth]{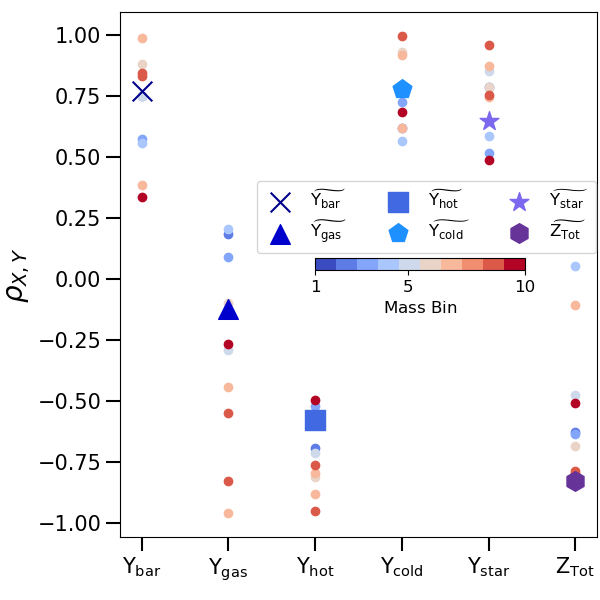}
\caption{Redshfit evolution from $0.2\lesssim z \lesssim 2$ of feedback, depletion parameters and metallicity and their correlations inside $R_{500, \mathrm c}$. (Left) Median values of depletion parameters, metallicity and feedback energy proxy as a function of redshift. The different lines (accordingly the legend on the bottom) represent the median values computed on the entire galaxy clusters sample in each redshift. (Right) Pearson correlation index of the redshift evolution computed between the feedback energy proxy and depletion parameters or metallicity. The coloured dots represent the median values computed in each mass bin (following the colour coding showed by the colourbar in the middle of the plot), while the dots identified by different markers (accordingly the legend in the middle of the plot) show the median values computed for the whole sample.}
\label{fig:feedback_r500}
\end{figure*}

Starting from the finding of \citet{Castro2021} and \citet{Ragagnin2022}, we investigate the role of AGN feedback in the time evolution of the depletion parameters. In the following, we do not include the energy feedback from stellar processes, because of its minor contribution with respect to the AGN feedback at the redshifts we are interested in \citep[see][and reference therein]{Ragagnin2022}. 
We define the feedback energy as the ratio between the mass accretion rate $\rm \dot{M}_{BH}$ and the thermal energy of galaxy clusters inside $R_{500, \mathrm c}$, 
\begin{equation} \label{eq:feedbackenergy}
    E_{feedback} = \frac{\dot{M}_{BH} \ c^2 \ \epsilon_{r} \ \epsilon_{f}}{(M_{gas,500c}/[\mu \ m_{p}] ) \times (T_{gas,500c}/[erg])}
\end{equation}
where $\dot{M}_{BH}$ is computed as the differences of the mean black hole mass between two consecutive time steps and $\epsilon_{r}$ and $\epsilon_{f}$ are the parameters proposed by \citet{2014MNRAS.442.2304H}, which takes into account the amount of feedback energy thermally coupled to the surrounding gas. For each redshift, we compute the median values of depletion parameters and metallicity for the entire galaxy clusters sample. 
The results are shown in the left panel of Fig.~\ref{fig:feedback_r500}. Here we observe that the feedback energy, rapidly decreases with the redshift, as already observed and discussed by \citet{Castro2021}. Moreover, we note that baryon, cold gas phase, and stellar depletion parameters show a decrease with redshift, whereas gas, hot gas phase, and metallicity have opposite behaviour. These trends suggest the presence of correlations between the feedback energy and the quantities analysed. In the right panel of Fig.~\ref{fig:feedback_r500}, we show the Pearson correlation indices computed between the time evolution of the feedback energy and the depletion parameters and metallicity. Focusing on the redshift range from $z\sim 2$ up to $z\sim 0.2$, we present the results both for the galaxy clusters divided in the same 10 mass bin adopted above and for the median values of the entire sample. As already observed, but now quantified by the Pearson correlation index, we note that the baryon depletion parameter, gas cold phase, and stellar ones, show a correlation with the energy feedback, while the gas depletion parameter, hot gas phase one, and metallicity show anti-correlated trends. In particular, we observe that in all the parameters analysed, except for the gas depletion parameter, the mass of the galaxy clusters does not change the correlation or anti-correlation trend. On the other hand, for the gas depletion parameter, we note a shift from correlation to anti-correlation with the increase in the mass of systems associated with the mass bins. This trend is associated with the relative impact of the different gas phases in different galaxy cluster sub-samples. Indeed, in less massive systems we note that the hot component is less dominant than in more massive galaxy clusters. This suggests that for less massive systems, the total gas depletion parameter is driven by the cold phase, which has a high level of correlation with feedback energy. Instead, for massive systems, the hot gas phase is completely dominant with respect to the cold one. The hot gas phase shows a high level of anti-correlation with the feedback energy, and this lead to an increase in the anti-correlation observed between the total gas depletion parameter and the feedback energy. In Tab.~\ref{tab:feedbackcorr} we report the Pearson correlation indices discussed above.  

\begin{table}[]
\centering
\begin{tabular}{c|ccc|}
&\multicolumn{3}{c|}{$\rho_{\rm X,Y}$} \\ \hline
&Less massive bin&Most massive bin&Entire sample \\ \hline \hline
$Y_{\rm bar}$&0.75&0.34&0.77 \\
$Y_{\rm gas}$&0.19&-0.27&-0.12 \\
$Y_{\rm hot}$&-0.69&-0.49&-0.58 \\
$Y_{\rm cold}$&0.62&0.68&0.78 \\
$Y_{\rm star}$&0.79&0.49&0.65 \\
$Z_{\rm Tot}$&-0.63&-0.51&-0.83 \\ \hline
\multicolumn{4}{c}{}\\ 
\end{tabular}
\caption{Pearson correlation index of the redshift evolution computed between the feedback energy proxy and depletion parameters or metallicity, inside $R_{500, \mathrm c}$, for the less massive sub-sample, the most massive one and the entire sample.}
\label{tab:feedbackcorr}
\end{table}

\begin{figure}
\includegraphics[width=0.49\textwidth]{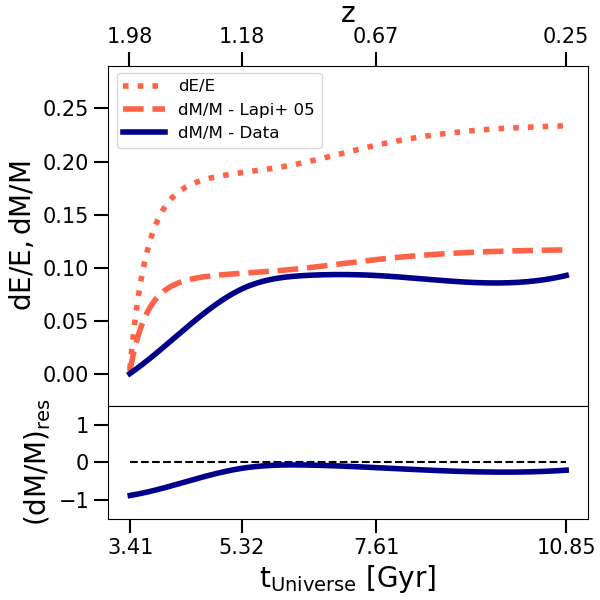}
\caption{Time evolution of feedback energy and mass depletion inside $R_{500, \mathrm c}$. (Top) The dotted line represents the time-integrated ratio between feedback energy and the system's thermal energy ("dE/E"), and the solid one shows the mass depletion computed as the difference between the baryon fraction in two consecutive time steps ("dM/M - Data"), the dashed line represents the expected mass depletion computed as $\rm \frac{1}{2}(dE/E)$, as proposed by \citet{Lapi2005} ("dM/M - Lapi+ 05"). The lines represent the median trends, computed over the whole galaxy clusters sample. (Bottom) The line shows the residuals of mass depletion, computed as $\rm ((dM/M)_{Data} - (dM/M)_{Lapi+05})/(dM/M)_{Lapi+05}$.}
\label{fig:dEdM_res}
\end{figure}

\citet{Lapi2005} proposed a model which relates the energy injected by an event of AGN activity and the fractional mass ejected by this event. In particular, the authors compute the energy introduced by an AGN event and the thermal energy of the hosting system. This quantity has the same meaning as our definition of feedback energy. \citet{Lapi2005} demonstrate that this energy ratio is related to the fractional mass ejected from the galaxy cluster environment. In detail, they demonstrate that $\Delta m/m \simeq \Delta E/2E$. To compare these findings with our analysis, we consider the fractional mass as the changing of baryon fraction over consecutive time steps and the time-integrate contribution of the feedback energy. In Fig.~\ref{fig:dEdM_res} we show the time evolution of time-integrate feedback energy (dE/E), the baryon fraction change (dM/M - Data) and the prediction on the fractional mass evolution given by the model of \citet{Lapi2005} (dM/M - Lapi+ 05). We note that the feedback energy rapidly increases around redshift $z=2$, while a flat behaviour is observed on lower redshift. On the other hand, the baryon fraction change increases more slowly than both the feedback energy and the prediction of the model, depending the latter only on the energy injected. The differences between the baryon fraction change we recover and the prediction of the model are related to the assumption adopted for the model. Indeed, the model proposed by \citet{Lapi2005} takes into account only heating phenomena, whereas at high redshift gravitational effects, such as accretion and mergers events, give not negligible contributions, making the model assumptions less effective. 

\begin{figure*}
\includegraphics[width=0.49\textwidth]{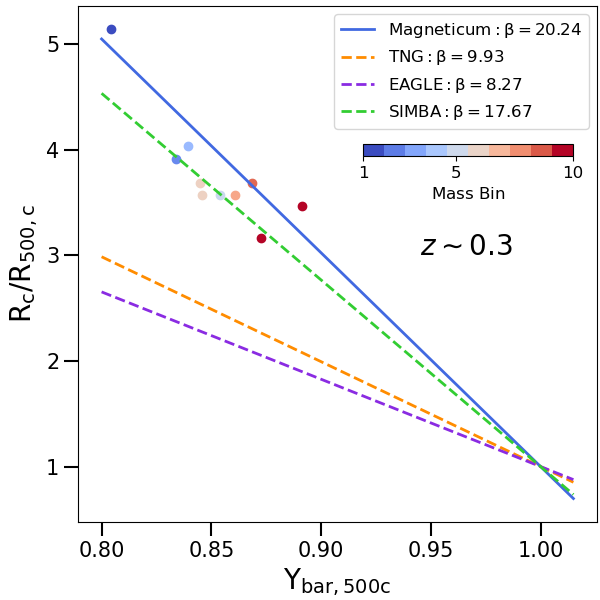}
\includegraphics[width=0.49\textwidth]{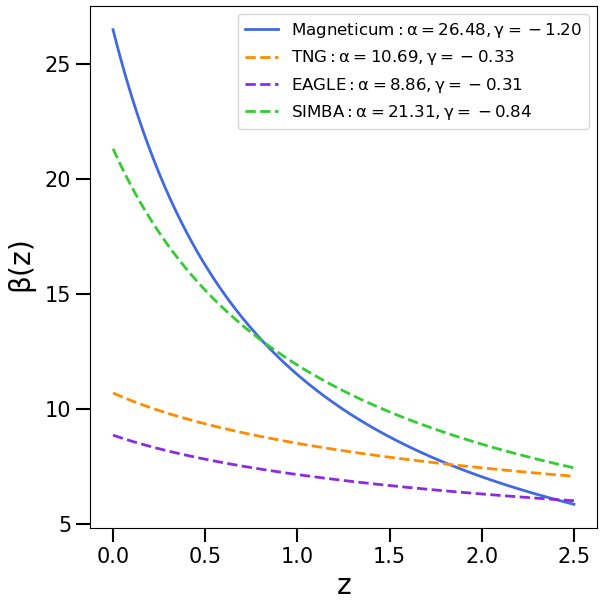}
\caption{Closure radius as function of baryon depletion factor within $R_{500, \mathrm c}$ and $\beta$ parameter as function of redshift. (Left) The dashed lines represent the results proposed by \citet{Ayromlou22} for Illustris-TNG (orange), EAGLE (purple), and SIMBA (green) simulations computed for redshift $z\sim 0.3$. The solid blue line shows the best fit relation between closure radius and baryon depletion factor, performed on our \textit{Magneticum} sample at redhsift $z\sim 0.3$. The coloured dots represent the median values of closure radius computed on the 10 mass bins (according to the colourbar on the top right corner of the plot) at the same redshift of $z\sim 0.3$. (Right) Redshift evolution of $\beta$ parameter. The dashed lines show the results proposed by \citet{Ayromlou22} for TNG (orange), EAGLE (purple), and SIMBA (green) simulations, whereas the solid blue line represents the findings we obtained as best fit on our \textit{Magneticum} sample.}
\label{fig:rc_beta}
\end{figure*}

Recently \citet{Ayromlou22} compare three different suites of numerical simulations, Illustris-TNG \citep{Pillepich2018,Nelson2018,Springel2018}, EAGLE \citep{Schaye2015,Crain2015} , and SIMBA \citep{Dave2019} to understand the evolution of baryon in halos with $M_{200,\mathrm c}$ masses between $10^8 M_{\odot}$ and $10^{15} M_{\odot}$, from halo's centre up to $30R_{200, \mathrm c}$. They demonstrate that baryon feedback mechanisms highly influence the baryon distribution, lowering the baryon budget within the halos and accumulating matter outside the virial radius of these systems. Moreover, they find that halos with different mass ranges are influenced by different feedback mechanisms. In particular, they show that for low-mass systems ($10^8\leq M_{200,\mathrm c}/M_{\odot}\leq10^{10}$) the main source of heating is given by the UV background, for intermediate mass halos ($10^{10}\leq M_{200,\mathrm c}/M_{\odot}\leq10^{12}$) stellar feedback becomes dominant, while for massive systems ($10^{12}\leq M_{200,\mathrm c}/M_{\odot}\leq10^{14}$) the main source of feedback is given by central AGN. Furthermore, they conclude that galaxy clusters with masses $M_{200,\mathrm c}/M_{\odot}\geq10^{14}$ are less affected by feedback phenomena instead by less massive objects. They also proposed a new characteristic scale, the closure radius $R_c$ that represents the radius at which all the baryons associated with a halo could be found. They define $R_c$ as:
\begin{equation}
    f_{\mathrm bar}(<R_c) = f_{\mathrm bar,cosmic} \pm \Delta f_{\mathrm bar,cosmic} 
\end{equation}
where $\Delta f_{\mathrm bar,cosmic}$ represents the observational uncertainty on the cosmic baryon fraction, which they assume to be $0.05$ \citep{Planck2016}. They find that the closure radius is closer to $R_{200,\mathrm c}$ in massive systems, while it tends to increasingly outer regions for objects in which the mass is gradually smaller. Moreover, they show that the position of the closure radius depends also on the model adopted by different simulations. Indeed, the simulations they used to give the different positions of closure radius for objects with the same masses. 
Starting from these findings, we compute the closure radius on our \textit{Magneticum} sample using the same definition proposed by \citet{Ayromlou22}. To compare our results with their finding, we also consider the universal relation they proposed, which relates to baryon fraction and closure radius:
\begin{equation}
    R_{c}/R_{500,\mathrm c} - 1 = \beta(z) \ (1 - f_{\mathrm bar}(<R_{500,\mathrm c})/f_{\mathrm bar,cosmic})
\end{equation}
with $\beta(z)$ is defined as:
\begin{equation}
    \beta(z) = \alpha (1+z)^\gamma
\end{equation}
where $\alpha$ and $\gamma$ are free parameters that we use to perform the fit on our simulations. 
In Fig.~\ref{fig:rc_beta} we show the comparison between our findings and the results proposed by \citet{Ayromlou22}. In the left panel, we present the closure radius, normalised to $R_{500, \mathrm c}$, as a function of baryon depletion factor $Y_{\mathrm bar}$ for the four different suites of numerical simulations at redshift $z\sim0.3$. The results for Illustris-TNG, EAGLE, and SIMBA simulations are derived from the best-fit parameters of $\alpha$ and $\gamma$ proposed by \citet{Ayromlou22}, while we perform the fitting procedure on our \textit{Magneticum} sample. We obtain values of $\alpha=26.48$ and $\gamma=-1.20$, which also determine the redshift evolution of $\beta(z)$ parameter proposed on the right panel of Fig.~\ref{fig:rc_beta}. As also discussed by \citet{Ayromlou22}, different models adopted for the treatment of AGN feedback highly influence the position of closure radius and cosmic time evolution of the $\beta$ parameter. Being the AGN model one of the main differences from the four different simulations analysed, this also suggests that the AGN feedback in galaxy clusters and groups represents the major source of energy responsible for the redistribution of baryon in the halo's environment. Moreover, the existence of the universal relation between closure radius and baryon fraction proposed by \citet{Ayromlou22} shows how will be crucial the next generation of X-rays observatories to observe clusters and groups peripheries and disentangle between different AGN feedback models. 

\section{Conclusions} \label{sec:conclusions}

In this work, we extend the analysis presented in \citet{Angelinelli22} by constraining the redshift evolution, up to $z=2.8$, of the distributions of baryons, gas, stars and metallicity up to large distances ($\sim 10R_{500,\mathrm c}$) from the centre of halos. We base our analysis on a large set of galaxy clusters from the \textit{Magneticum} simulation, and investigate the mass-dependent effects, by dividing further our sample of $150$ clusters into 10 mass bins, and separately studying the evolving spatial distribution of the multi-phase baryonic across the sample. 


Our main findings can be so summarised:
\begin{itemize}
    \item In the central region of galaxy clusters ($r<R_{500,\mathrm c}$) the baryon fraction shows a general decrease with the redshift, with a decrease for less massive objects which is $\sim4$ times larger than in massive ones (see Fig.~\ref{fig:ybar_r500} and Tab.~\ref{tab:fixradii_redshift}). The gas depletion parameter we recover inside $R_{500,\mathrm c}$ is an agreement with observations of low redshift halos \citep[][]{Eckert21,Akino22} (see Fig.~\ref{fig:ygasyhot_r500}). At high redshift ($z>1.2$) instead, the contribution given by the cold gas phase ($kT < 0.1$ keV) is non-negligible. For instance, at $z\sim2.8$ the cold component accounts for $\sim20\div28\%$ of the total gas, depending on the sub-sample mass. This leads to an increase in the total amount of gas, not well matched by the best fit suggested from low-z observations. 
    
    \item We measure a clear redshift evolution of the simulated radial behaviour of baryon depletion parameter, up to $10R_{500,\mathrm c}$ (see Fig.~\ref{fig:ybar_radial} and Tab.~\ref{tab:fixradii_redshift}). In agreement with our previous work {\it PaperI}, the "closed-box" assumption is valid, at lower redshifts, only in massive galaxy clusters and on radii greater than $5R_{500,\mathrm c}$. On the other hand, the same condition is verified at higher redshifts independently on the mass and for $r\geq 3R_{500,\mathrm c}$. 
    The radial trend of the gas depletion parameter shows that the differences between the less and the most massive objects increase with the cosmic time (Fig.~\ref{fig:ygas_radial}). Even if the slope of the profiles is steeper for less massive objects, at all redshifts, the scatter in the central regions in the most massive sub-sample is half of the same quantity for less massive galaxy clusters. At larger radii ($r\geq2R_{500,\mathrm c}$), instead, the trend and absolute values of the gas depletion parameter are similar, regardless of mass or redshift. 
    The hot gas phase depletion parameter increases radially with the time at all masses (see Fig.~\ref{fig:yhot_radial}). At high redshift, the hot gas phase is not enough to completely describe the gas depletion parameter. This means that the cold and X-ray unobserved gas components cannot be neglected to close the cosmic baryon budget of high redshift ($z>1.2$) halos. 
    
    \item Previous works from \citet{Castro2021} and \citet{Ragagnin2022} show that AGN feedback phenomena affect the evolution of baryon and gas fraction in galaxy clusters environment.  We additionally studied the relation between the feedback energy (Eq.~\ref{eq:feedbackenergy}) and the depletion parameter (see Fig.~\ref{fig:feedback_r500} and Tab.~\ref{tab:feedbackcorr}), found a high level of correlation with the evolution of the cold gas phase and of the stellar depletion parameter. On the other hand, the gas, the hot gas phase, and the gas metallicity are anti-correlated with the evolving feedback energy. The gas depletion parameter is the only parameter that also shows a dependency on the correlation with the mass of the galaxy clusters; this is expected because in less massive objects the contribution of cold gas is not negligible as instead for most massive systems. This trend appears consistent with early theoretical work by \citet{Lapi2005} (see Fig.~\ref{fig:dEdM_res}), suggesting that indeed the low redshift evolution of the gas depletion parameter is mainly driven by AGN feedback. The role of AGN feedback in numerical simulations is also discussed in \citet{Ayromlou22}. Studying Illustris-TNG, EAGLE, and SIMBA simulations, they propose a new characteristic scale, the closure radius $R_c$, and a universal relation between $R_c$ and $f_{\rm bar}$. After computing $R_c$ for our sample, we test the universal relation proposed by \citet{Ayromlou22} on \textit{Magneticum} simulation (see Fig.~\ref{fig:rc_beta}). We confirm that different models of AGN feedback implemented in different suites of numerical simulations give different findings of $R_c$ and also determine different best-fit parameters for the universal relation between $R_c$ and $f_{bar}$. These results suggest the need for further observational investigation to find the model of AGN feedback that better reproduces the observational constraints. On the other hand, the analysis of gas metallicity and stellar depletion parameter (see Fig.~\ref{fig:ztot_radial} and Fig.~\ref{fig:ystar_radial}) suggests that early enrichment scenario \citep[see][]{biffi17,biffi18} is likely to account for the aforementioned trend with mass and redshift. 
    In the central regions of galaxy clusters, the redshift evolution of gas metallicity accounts for less than 20\% in less massive objects and less than 10\% in the most massive ones. Moreover, the stellar depletion parameter decrease by a factor of 2, independently from the mass of the galaxy clusters. These trends suggest that recent star-formation processes give negligible contributions to the enrichment of the gas metallicity.
    
    . 
    \item Finally, we extended the functional form proposed in {\it PaperI} taking into account the dependencies on the radius, mass, and redshift of the baryon, gas, and hot gas phase depletion parameters. 
    The functional form is described by the formula:
    \begin{equation} \label{eq:functionalform}
    Y_{i} = \alpha \cdot w^\beta \cdot x^{\gamma + \delta*w +\epsilon*(1+z)}
    \end{equation}
    where $w = M_{500, \mathrm c} / 5\cdot10^{14} \ h^{-1}M_{\odot}$, $x = r /R_{500, \mathrm c}$, $z$ is the redshift, while $\alpha$, $\beta$, $\gamma$, $\delta$ and $\epsilon$ are the free parameters. 
    Our best-fit values (Tab.~\ref{tab:fitdepletion}) are able to provide a description within 2\% for the baryon depletion parameter and within 3\% for gas and hot gas phase ones (see Fig.~\ref{fig:ybarygasyhot_funct}).  
\end{itemize}


These relations could be used as a proxy in current and future X-rays observations to provide useful constraints to test the different AGN feedback models used in different suites of numerical simulations.

\section*{Acknowledgements}

M.A. acknowledges the financial support from "Borsa Marco Polo – II tornata BIR 2021".
S.E. acknowledges the financial contribution from the contracts ASI-INAF Athena 2019-27-HH.0,
``Attivit\`a di Studio per la comunit\`a scientifica di Astrofisica delle Alte Energie e Fisica Astroparticellare''
(Accordo Attuativo ASI-INAF n. 2017-14-H.0), INAF mainstream project 1.05.01.86.10, and
from the European Union’s Horizon 2020 Programme under the AHEAD2020 project (grant agreement n. 871158).
F.V. acknowledges financial support from the Horizon 2020 program under the ERC Starting Grant MAGCOW, no. 714196. 
KD acknowledges support by the COMPLEX project from the European Research Council (ERC) under the European Union’s Horizon 2020 research and innovation program grant agreement ERC-2019-AdG 882679.
The Magneticum Simulations were carried out at the Leibniz Supercomputer Center (LRZ) under the project pr83li. 

\bibliographystyle{aa}
\bibliography{biblio}

\begin{appendix}

\section{Depletion parameters estimations and fitting results}

In Tab.~\ref{tab:fixradii_redshift} we report the details on the values of the depletion parameters under investigations in the less and the most massive bins and at four radii ($R_{500,\mathrm c}, 3R_{500,\mathrm c}, 5R_{500,\mathrm c}$ and $10R_{500,\mathrm c}$) and for four different redshifts (2.79, 1.71, 0.67 and 0.25).

\begin{figure*}
\includegraphics[width=0.33\textwidth]{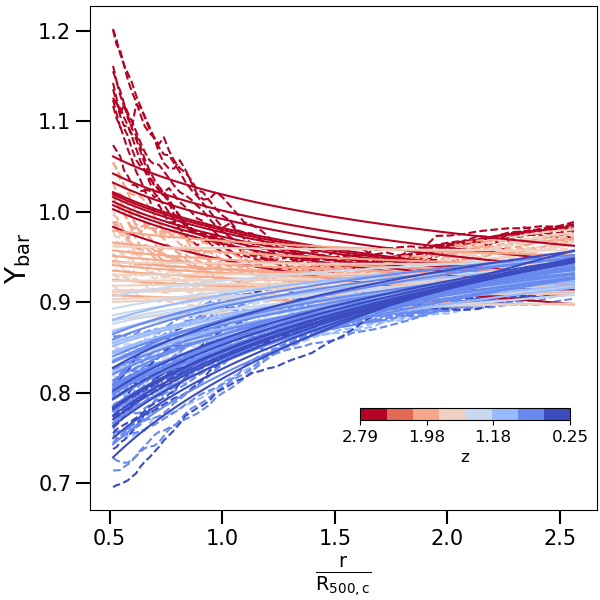}
\includegraphics[width=0.33\textwidth]{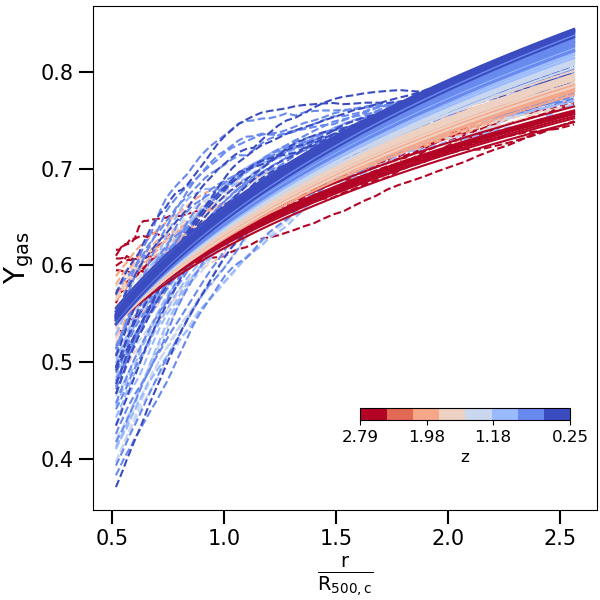}
\includegraphics[width=0.33\textwidth]{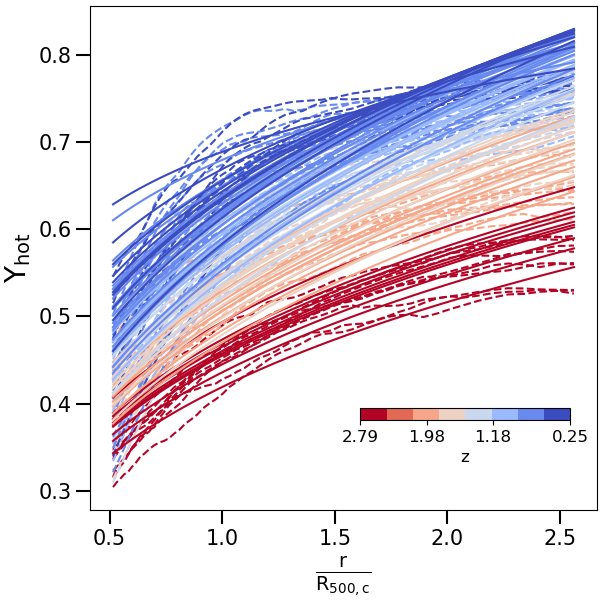}
\caption{Radial profiles of baryon (left), gas (centre) and hot gas phase (right) depletion, from $0.5R_{500,\mathrm c}$ up to $2.5R_{500,\mathrm c}$. The dashed lines represent the median profiles of each mass bin at each redshift, according to the colourbar in the bottom right corner. The solid lines are the fit performed according to functional form Eq.~\ref{eq:functionalform}, with the same colour scale of median profiles.}
\label{fig:ybarygasyhot_funct}
\end{figure*}

We provide within Tab.~\ref{tab:fitdepletion} (see also Fig.~\ref{fig:ybarygasyhot_funct}) the best-fit parameters obtained from the fitting function in Eq.~\ref{eq:functionalform}.
Following our previous work {\it PaperI}, we limit our fitting produce in a radial range comparable with the one of present and near future X-rays observations, focusing in the central regions of our analysis, from $0.5R_{500,\mathrm c}$ to $2.5R_{500,\mathrm c}$.
The functional form is able to well reproduce the behaviour of all the depletion parameters analysed, as shown by the $\Tilde{\chi}^2$ and values of the median and the maximum deviation of the model from the data ($\Tilde{e}$ and $e_{max}$, respectively). However, as already observed in {\it PaperI}, also in this case, for the $Y_{\rm gas}$ and $Y_{\rm hot}$ we note that our fitting procedure gives less strong results than for $Y_{\rm bar}$ case. We use the dispersion around the mean profile as the weight to evaluate $\chi^2$.

\begin{table*}
\centering
\begin{tabular}{c|c|cccc|cccc|}
&&\multicolumn{4}{c|}{Less massive sub-sample}&\multicolumn{4}{c|}{Most massive sub-sample} \\ \hline
z&&$R_{500,\mathrm c}$&$3R_{500,\mathrm c}$&$5R_{500,\mathrm c}$&$10R_{500,\mathrm c}$ &$R_{500,\mathrm c}$&$3R_{500,\mathrm c}$&$5R_{500,\mathrm c}$&$10R_{500,\mathrm c}$\\ \hline \hline
\multirow{7}{*}{2.79}&&\multicolumn{4}{c|}{$5.8\times 10^{12} \leq M_{500,\mathrm c}/M_{\odot}\leq 9.7\times 10^{12}$}&\multicolumn{4}{c|}{$2.5\times 10^{13} \leq M_{500,\mathrm c}/M_{\odot}\leq 5.0\times 10^{13}$} \\
&$Y_{\rm bar}$&$0.99^{+0.02}_{-0.05}$&$0.99^{+0.01}_{-0.02}$&$1.0^{+0.01}_{-0.01}$&$1.0^{+0.01}_{-0.01}$&$0.96^{+0.01}_{-0.02}$&$0.99^{+0.01}_{-0.01}$&$1.0^{+0.01}_{-0.01}$&$1.00^{+0.01}_{-0.01}$ \\
&$Y_{\rm gas}$&$0.64^{+0.02}_{-0.06}$&$0.78^{+0.02}_{-0.03}$&$0.82^{+0.02}_{-0.01}$&$0.88^{+0.01}_{-0.01}$&$0.63^{+0.02}_{-0.03}$&$0.79^{+0.01}_{-0.02}$&$0.84^{+0.01}_{-0.01}$&$0.90^{+0.01}_{-0.01}$ \\
&$Y_{\rm hot}$&$0.45^{+0.02}_{-0.04}$&$0.55^{+0.02}_{-0.06}$&$0.44^{+0.05}_{-0.03}$&$0.27^{+0.04}_{-0.05}$&$0.50^{+0.02}_{-0.06}$&$0.57^{+0.04}_{-0.02}$&$0.46^{+0.04}_{-0.02}$&$0.27^{+0.02}_{-0.01}$ \\
&$Y_{\rm cold}$&$0.18^{+0.05}_{-0.03}$&$0.25^{+0.03}_{-0.06}$&$0.38^{+0.04}_{-0.06}$&$0.62^{+0.06}_{-0.05}$&$0.13^{+0.04}_{-0.01}$&$0.22^{+0.03}_{-0.04}$&$0.38^{+0.02}_{-0.04}$&$0.63^{+0.01}_{-0.03}$ \\
&$Y_{\rm star}$&$0.33^{+0.09}_{-0.04}$&$0.20^{+0.03}_{-0.01}$&$0.18^{+0.01}_{-0.02}$&$0.11^{+0.01}_{-0.01}$&$0.33^{+0.07}_{-0.03}$&$0.20^{+0.02}_{-0.01}$&$0.16^{+0.02}_{-0.01}$&$0.10^{+0.01}_{-0.01}$ \\
&$Z_{\rm tot}$&$0.05^{+0.04}_{-0.03}$&$0.03^{+0.27}_{-0.02}$&$0.07^{+0.20}_{-0.05}$&$0.02^{+0.27}_{-0.02}$&$0.09^{+0.11}_{-0.03}$&$0.03^{+0.09}_{-0.02}$&$0.03^{+0.10}_{-0.02}$&$0.02^{+0.04}_{-0.01}$ \\ \hline
\multirow{6}{*}{1.71}&&\multicolumn{4}{c|}{$1.2\times 10^{13} \leq M_{500,\mathrm c}/M_{\odot}\leq 3.4\times 10^{13}$}&\multicolumn{4}{c|}{$6.0\times 10^{13} \leq M_{500,\mathrm c}/M_{\odot}\leq 1.5\times 10^{14}$} \\
&$Y_{\rm bar}$&$0.91^{+0.03}_{-0.08}$&$0.97^{+0.02}_{-0.03}$&$0.99^{+0.01}_{-0.01}$&$1.00^{+0.01}_{-0.01}$&$0.91^{+0.03}_{-0.01}$&$0.98^{+0.01}_{-0.01}$&$1.00^{+0.01}_{-0.01}$&$1.00^{+0.01}_{-0.01}$ \\
&$Y_{\rm gas}$&$0.64^{+0.02}_{-0.09}$&$0.80^{+0.02}_{-0.01}$&$0.85^{+0.02}_{-0.01}$&$0.89^{+0.01}_{-0.01}$&$0.68^{+0.02}_{-0.01}$&$0.82^{+0.01}_{-0.01}$&$0.86^{+0.01}_{-0.01}$&$0.90^{+0.01}_{-0.01}$ \\
&$Y_{\rm hot}$&$0.53^{+0.03}_{-0.08}$&$0.67^{+0.03}_{-0.01}$&$0.62^{+0.03}_{-0.02}$&$0.41^{+0.03}_{-0.05}$&$0.61^{+0.02}_{-0.01}$&$0.72^{+0.02}_{-0.02}$&$0.65^{+0.02}_{-0.04}$&$0.43^{+0.04}_{-0.05}$ \\
&$Y_{\rm cold}$&$0.09^{+0.03}_{-0.03}$&$0.12^{+0.03}_{-0.02}$&$0.23^{+0.04}_{-0.03}$&$0.49^{+0.04}_{-0.05}$&$0.07^{+0.01}_{-0.01}$&$0.10^{+0.02}_{-0.02}$&$0.21^{+0.05}_{-0.02}$&$0.47^{+0.06}_{-0.05}$ \\
&$Y_{\rm star}$&$0.26^{+0.04}_{-0.03}$&$0.17^{+0.01}_{-0.02}$&$0.14^{+0.01}_{-0.01}$&$0.10^{+0.01}_{-0.01}$&$0.23^{+0.02}_{-0.01}$&$0.16^{+0.01}_{-0.01}$&$0.13^{+0.01}_{-0.01}$&$0.10^{+0.01}_{-0.01}$ \\
&$Z_{\rm tot}$&$0.22^{+0.17}_{-0.06}$&$0.17^{+0.17}_{-0.10}$&$0.15^{+0.08}_{-0.10}$&$0.07^{+0.25}_{-0.02}$&$0.19^{+0.03}_{-0.07}$&$0.09^{+0.11}_{-0.04}$&$0.11^{+0.08}_{-0.04}$&$0.07^{+0.14}_{-0.04}$ \\ \hline
\multirow{6}{*}{0.67}&&\multicolumn{4}{c|}{$6.0\times 10^{13} \leq M_{500,\mathrm c}/M_{\odot}\leq 1.0\times 10^{14}$}&\multicolumn{4}{c|}{$2.3\times 10^{14} \leq M_{500,\mathrm c}/M_{\odot}\leq 4.6\times 10^{14}$} \\
&$Y_{\rm bar}$&$0.81^{+0.05}_{-0.01}$&$0.93^{+0.01}_{-0.01}$&$0.98^{+0.01}_{-0.02}$&$1.00^{+0.01}_{-0.01}$&$0.92^{+0.02}_{-0.02}$&$0.96^{+0.02}_{-0.02}$&$0.99^{+0.01}_{-0.01}$&$1.00^{+0.01}_{-0.01}$ \\
&$Y_{\rm gas}$&$0.60^{+0.07}_{-0.01}$&$0.80^{+0.01}_{-0.02}$&$0.87^{+0.01}_{-0.02}$&$0.90^{+0.01}_{-0.01}$&$0.73^{+0.01}_{-0.02}$&$0.82^{+0.01}_{-0.01}$&$0.87^{+0.01}_{-0.01}$&$0.90^{+0.01}_{-0.01}$ \\
&$Y_{\rm hot}$&$0.56^{+0.07}_{-0.01}$&$0.76^{+0.01}_{-0.02}$&$0.77^{+0.01}_{-0.02}$&$0.62^{+0.01}_{-0.05}$&$0.70^{+0.02}_{-0.02}$&$0.78^{+0.01}_{-0.01}$&$0.78^{+0.01}_{-0.03}$&$0.63^{+0.06}_{-0.07}$ \\
&$Y_{\rm cold}$&$0.04^{+0.01}_{-0.01}$&$0.04^{+0.01}_{-0.01}$&$0.10^{+0.02}_{-0.01}$&$0.28^{+0.05}_{-0.01}$&$0.03^{+0.01}_{-0.01}$&$0.04^{+0.01}_{-0.01}$&$0.09^{+0.03}_{-0.03}$&$0.27^{+0.07}_{-0.07}$ \\
&$Y_{\rm star}$&$0.21^{+0.01}_{-0.02}$&$0.13^{+0.01}_{-0.01}$&$0.12^{+0.01}_{-0.01}$&$0.09^{+0.01}_{-0.01}$&$0.19^{+0.01}_{-0.01}$&$0.13^{+0.01}_{-0.01}$&$0.12^{+0.01}_{-0.01}$&$0.09^{+0.01}_{-0.01}$ \\
&$Z_{\rm tot}$&$0.34^{+0.03}_{-0.04}$&$0.28^{+0.14}_{-0.04}$&$0.17^{+0.08}_{-0.03}$&$0.14^{+0.05}_{-0.04}$&$0.34^{+0.08}_{-0.06}$&$0.23^{+0.10}_{-0.04}$&$0.18^{+0.11}_{-0.05}$&$0.21^{+0.09}_{-0.09}$ \\ \hline
\multirow{6}{*}{0.25}&&\multicolumn{4}{c|}{$6.0\times 10^{13} \leq M_{500,\mathrm c}/M_{\odot}\leq 1.9\times 10^{14}$}&\multicolumn{4}{c|}{$4.6\times 10^{14} \leq M_{500,\mathrm c}/M_{\odot}\leq 7.5\times 10^{14}$} \\
&$Y_{\rm bar}$&$0.84^{+0.02}_{-0.01}$&$0.93^{+0.02}_{-0.01}$&$0.98^{+0.01}_{-0.02}$&$1.00^{+0.01}_{-0.01}$&$0.92^{+0.01}_{-0.04}$&$0.93^{+0.02}_{-0.01}$&$0.99^{+0.01}_{-0.01}$&$1.00^{+0.01}_{-0.01}$ \\
&$Y_{\rm gas}$&$0.67^{+0.01}_{-0.04}$&$0.81^{+0.02}_{-0.01}$&$0.88^{+0.01}_{-0.02}$&$0.90^{+0.01}_{-0.01}$&$0.75^{+0.03}_{-0.04}$&$0.81^{+0.03}_{-0.01}$&$0.87^{+0.01}_{-0.01}$&$0.90^{+0.01}_{-0.01}$ \\
&$Y_{\rm hot}$&$0.64^{+0.01}_{-0.04}$&$0.78^{+0.02}_{-0.01}$&$0.80^{+0.02}_{-0.01}$&$0.68^{+0.07}_{-0.04}$&$0.73^{+0.03}_{-0.04}$&$0.79^{+0.02}_{-0.01}$&$0.82^{+0.01}_{-0.01}$&$0.70^{+0.02}_{-0.03}$ \\
&$Y_{\rm cold}$&$0.02^{+0.01}_{-0.01}$&$0.03^{+0.01}_{-0.01}$&$0.07^{+0.01}_{-0.01}$&$0.22^{+0.04}_{-0.08}$&$0.02^{+0.01}_{-0.01}$&$0.02^{+0.01}_{-0.01}$&$0.05^{+0.02}_{-0.01}$&$0.21^{+0.03}_{-0.02}$ \\
&$Y_{\rm star}$&$0.18^{+0.02}_{-0.01}$&$0.12^{+0.01}_{-0.01}$&$0.11^{+0.01}_{-0.01}$&$0.09^{+0.01}_{-0.01}$&$0.17^{+0.01}_{-0.02}$&$0.12^{+0.01}_{-0.01}$&$0.11^{+0.01}_{-0.01}$&$0.09^{+0.01}_{-0.01}$ \\
&$Z_{\rm tot}$&$0.32^{+0.08}_{-0.06}$&$0.25^{+0.09}_{-0.06}$&$0.25^{+0.09}_{-0.09}$&$0.19^{+0.06}_{-0.05}$&$0.34^{+0.07}_{-0.06}$&$0.23^{+0.06}_{-0.03}$&$0.21^{+0.14}_{-0.05}$&$0.18^{+0.05}_{-0.05}$ \\ \hline
\multicolumn{10}{c}{}\\
\end{tabular}
\caption{Baryons ($Y_{\rm bar}$), gas ($Y_{\rm gas}$), hot gas phase ($Y_{\rm hot}$), cold gas phase ($Y_{\rm cold}$) and stellar ($Y_{\rm star}$) depletion parameters and gas metallicity ($Z_{tot}$) computed at four different radii ($1, 3, 5$ and $10$ times $R_{500,\mathrm c}$) for the less massive sub-sample (left side) and the most massive one (right side), in each of the four different redshifts ($2.79, 1.70, 0.67$ and $0.25$). Errors are given as 16th and 84th distributions percentiles. Note that the depletion parameters are computed within the given radii, differently from the metallicity values which are given within a spherical shell (considering the same reference radii).}
\label{tab:fixradii_redshift}
\end{table*}

\begin{table*}
\centering
\hspace{5pt}
\begin{tabular}{c|ccc|}
&$Y_{\rm bar}$&$Y_{\rm gas}$&$Y_{\rm hot}$ \\ \hline 
$\alpha$&$0.821\pm0.001$&$0.660\pm0.001$&$0.680\pm0.001$ \\ \hline
$\beta$&$-0.051\pm0.001$&$0.015\pm0.001$&$0.115\pm0.001$ \\ \hline
$\gamma$&$0.129\pm0.006$&$0.315\pm0.008$&$0.482\pm0.008$ \\ \hline
$\delta$&$0.093\pm0.007$&$-0.015\pm0.008$&$-0.254\pm0.008$ \\ \hline
$\epsilon$&$-0.050\pm0.001$&$-0.029\pm0.002$&$-0.047\pm0.003$ \\ \hline
$\Tilde{\chi}^2$&$1.62$&$1.03$&$0.83$ \\ \hline
$\Tilde{e}$&$2\%$&$3\%$&$3\%$ \\ \hline
$e_{max}$&$17\%$&$32\%$&$28\%$ \\ \hline
\multicolumn{4}{c}{}\\
\end{tabular}
\caption{Best-fit parameters and related standard errors, for the functional form Eq.~\ref{eq:functionalform} fitted on $Y_{\rm bar}$, $Y_{\rm gas}$ and $Y_{\rm hot}$. The values of the reduced $\chi^2$ ($\Tilde{\chi}^2$), the median and the maximum deviation of the model from the data ($\Tilde{e}$ and $e_{max}$, respectively) are also quoted.}
\label{tab:fitdepletion}
\end{table*}

\end{appendix}

\end{document}